\documentclass{sig-alternate-2013}

\usepackage[show]{notes-alt}
\usepackage{tabularx}
\usepackage{tabulary}
\usepackage{booktabs}
\usepackage{datetime}
\usepackage[show]{notes-alt}
\usepackage{subfigure}
\usepackage[square,numbers]{natbib}

\usepackage{epstopdf}
\usepackage{float}
\usepackage{url}
\usepackage{hyperref}
\usepackage[all]{hypcap}
\usepackage{tabto}
\usepackage{amsmath}
\usepackage{enumitem}
\usepackage[ruled, linesnumbered]{algorithm2e}
\usepackage{rotating}
\usepackage{color, colortbl}
\usepackage{soul}

\newcommand{\hide}[1]{}

\permission{Copyright is held by the author/owner(s).}
\conferenceinfo{WWW'16 Companion,}{April 11--15, 2016, Montr\'eal, Qu\'ebec, Canada.} 
\copyrightetc{ACM \the\acmcopyr}
\crdata{978-1-4503-4144-8/16/04. \\
http://dx.doi.org/10.1145/2872518.2890555}

\begin{document}

\title{Tempas: Temporal Archive Search Based on Tags\thanks{This work is partly funded by the European Research Council under ALEXANDRIA (ERC 339233)}
}

\numberofauthors{2}
\author{
\alignauthor
Helge Holzmann\\
       \affaddr{L3S Research Center}\\
       \affaddr{Appelstr. 9a}\\
       \affaddr{30167 Hanover, Germany}\\
       \email{holzmann@L3S.de}
\alignauthor
Avishek Anand\\
       \affaddr{L3S Research Center}\\
       \affaddr{Appelstr. 9a}\\
       \affaddr{30167 Hanover, Germany}\\
       \email{anand@L3S.de}
}

\maketitle

\begin{abstract}
Limited search and access patterns over Web archives have been well documented. One of the key reasons is the lack of understanding of the user access patterns over such collections, which in turn is attributed to the lack of effective search interfaces. Current search interfaces for Web archives are (a) either purely navigational or (b) have sub-optimal search experience due to ineffective retrieval models or query modeling. We identify that external longitudinal resources, such as social bookmarking data, are crucial sources to identify important and popular websites in the past. To this extent we present \textsf{Tempas}, a tag-based temporal search engine for Web archives.

Websites are posted at specific times of interest on several external platforms, such as bookmarking sites like \emph{Delicious}. Attached tags not only act as relevant descriptors useful for retrieval, but also encode the time of relevance. With \textsf{Tempas} we tackle the challenge of temporally searching a Web archive by indexing tags and time. We allow temporal selections for search terms, rank documents based on their popularity and also provide meaningful query recommendations by exploiting tag-tag and tag-document co-occurrence statistics in arbitrary time windows. Finally, \textsf{Tempas} operates as a fairly non-invasive indexing framework. By not dealing with contents from the actual Web archive it constitutes an attractive and low-overhead approach for quick access into Web archives.

\end{abstract}

%
%


\newpage

\section{Introduction}

Web archiving initiatives such as the \emph{Internet Archive}\footnote{\url{http://www.archive.org}} and the \emph{Internet Memory Foundation}\footnote{\url{http://www.internetmemory.org}} have been involved in periodically archiving websites for over 17 years with collection sizes amounting to several hundreds of terabytes. Longitudinal collections present many opportunities for various kinds of historical analyses~\cite{Schreibman:2008:CDH:1796118}, Cliodynamics, cultural analyses and Culturomics~\cite{Michel:ys}, as well as analytics for computational journalism~\cite{CohenLYY:cidr11}. 
Long-term preservation is the first step, but the true potential of these collections can only be realized by enabling \emph{effective search} and \emph{exploration} over such collections. 

Unfortunately, search over Web archives have been limited. Companies and organizations maintaining Web archives either provide only very rudimentary search interfaces or pure URL-lookup services like the \emph{Wayback Machine}\footnote{\url{http://archive.org/web/}}. For that reason, access methods for Web archives are limited, also because usage patterns on Web archives as a corpus of study are not very well understood. This results in a lack of training data for user intents and information needs. Due to the size of data in those archives and their temporal aspect, out-of-the-box search infrastructures with full-text indexes are resource and computationally expensive and largely do not fit the needs. Since challenges like temporal ranking, link analysis and diversification are not solved yet \citep{Campos2014}, it would be hard for the user to identify authoritative documents and effectively rank retrieved results.

In view of these issues we propose an alternative search approach, which exploits external data sources as a proxy for popular and historically relevant websites, instead of trying to compute those metrics on internal features of the archived websites. For that purpose we built \textsf{Tempas} on the social bookmarking platform \emph{Delicious}\cite{socialbm0311}. Generally, we explore the idea of taking advantage of social media metadata about websites in the archive as cues for website importance. It is not uncommon for commercial search engines to cross reference social media feedback in designing features for the same reason. One distinct advantage of such data is direct human endorsement of websites that they find interesting. A second and even more important aspect for the perspective of a temporal search engine is its temporal annotation. Metadata from external sources like social networks is typically timestamped, which proves to be a useful asset in identifying temporal importance of websites.




\emph{Delicious} as a popular social bookmarking platform is one of the social media services that works in this fashion. It is intended exactly for the case of posting links together with metadata which describes a website in order to save the information and to share with others. On \emph{Delicious} tags are used as meta information to describe a website at the time of post. Tags are single terms, for instance topics and subtopics, which together describe a the theme of a website. Applicability of tags for search tasks has been previously explored in \citep{BischoffF08,Paiu09}. However, they failed to take the temporal aspects of tagging into account and thus fall short for longitudinal collections like Web archives. With \textsf{Tempas} we present a search engine incorporating tags from \textit{Delicious} in order to enable richer search capabilities on archived websites than currently available. Our demonstration implements this idea in an interactive Web application that allows users to search and explore Web archives using tags as proxies.


\section{Demo Scenario}

\begin{figure*}[t!]
\centering
\includegraphics[width=\textwidth]{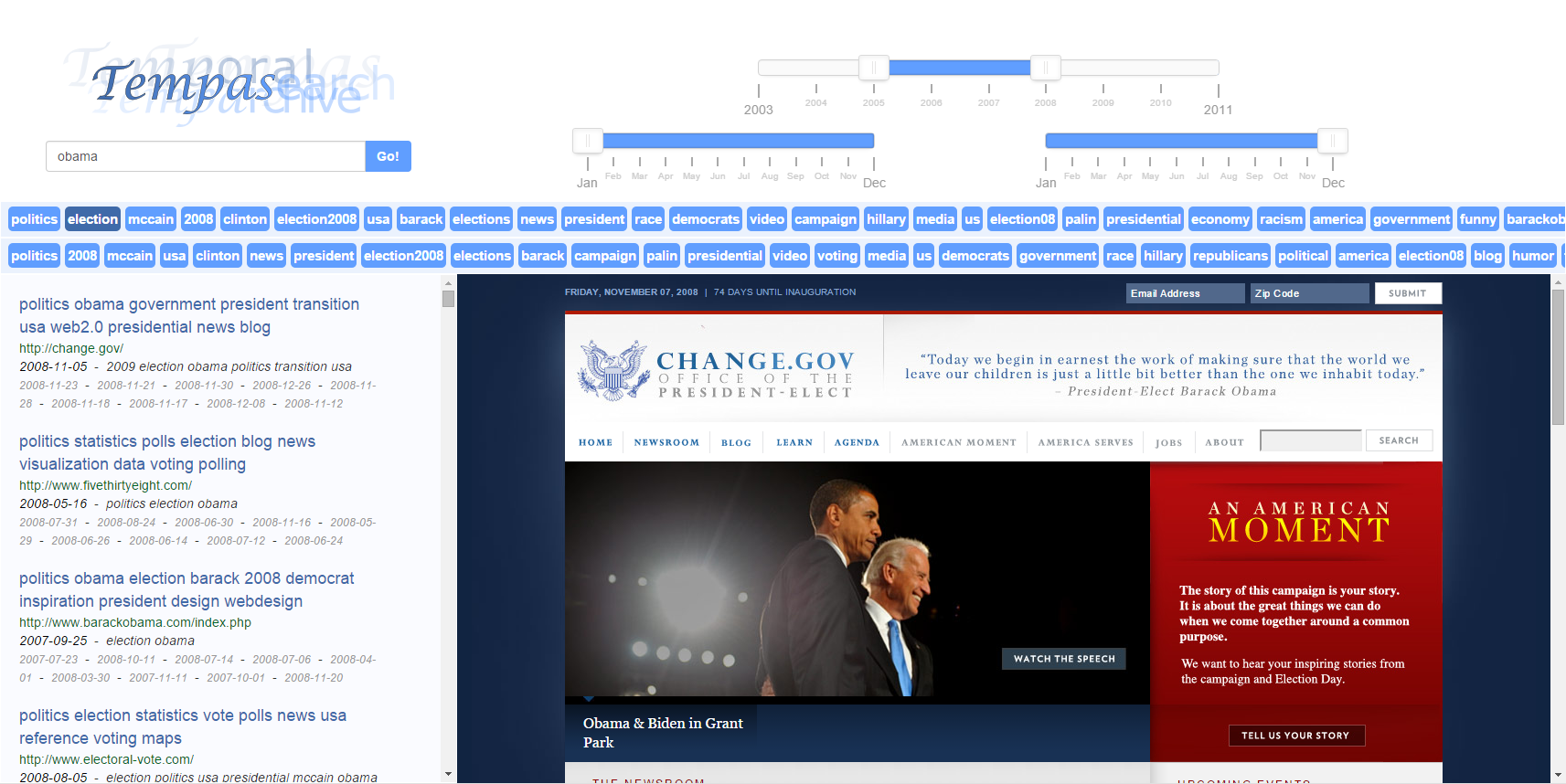}
\caption{\textsf{Tempas} showing search results for \textit{obama} and \textit{election} between Jan 2005 and Dec 2008.}
\label{fig:screenshot}
\end{figure*}

A typical scenario that \textsf{Tempas} can be used for is to find event related websites in a Web archive, i.e., websites related to a certain topic in a given time period. This can be important for researcher who want to study a topic from the past with the Web archive as their scholarly source. Today, those websites might not be relevant for that topic anymore or do not even exist anymore. For that reason, the website will not be available in the index of a current search engine or cannot be discovered by a user using the same keywords. Thus, it cannot be viewed in the Wayback Machine since that would require to know the URL.

An example of such a scenario is the election campaign website of Barack Obama for the US presidential election in 2008: \textit{change.gov}. Today, the website shows an image stating the transition has ended and the new administration has begun\footnote{\url{http://www.change.gov}, visited: 22/12/2015}. Also it is not among the top search results on Google anymore. A query for \textit{obama election 2008} primarily yields more current websites reporting about the election. A researcher who is interested in reproducing the campaign might however be more interested in original content from that time. Also regular users who just want to revisit the pre-election promises to compare with the achievement of the elected government will need to look at the original websites from back then.

A query for \textit{obama} on \textsf{Tempas} during the time when he was senator of Illinois from 2005 to 2008 before he became president, looks as shown in Figure~\ref{fig:screenshot}. On the first suggestion bar it lists those terms that were most relevant to the query during the selected time, which are frequently co-occurring tags of the issued query. Of course, one of the top ranked tags here is the election, which can be selected to refine the search results and focus on this particular sub-topic. This opens up a second suggestion bar which is slightly more aligned towards the election and reranks the tags according to their co-occurrence with both query tags \textit{obama} and \textit{election} during the selected time period. The results shown in the left panel are those websites which were most relevant for the users of delicious with respect to the given query terms and time. Besides Barack Obama's homepage and a website on statistics about the election on the second and third rank, the first rank is actually his election campaign website \textit{change.gov}.

To get an impression what is behind those websites, similar to search results on Google or other search engines, every result includes a title. On \textsf{Tempas} this comprises the most related tags during the time period of the query. For the desired election campaign website, these tags describe it as a political website of Obama and his government, which includes news and blog articles. This description does not necessarily correspond to the content of the websites, but instead represents a temporal view on the websites by its visitors.

Up to this point, all information has been compiled purely based our external source delicious, without deriving data from the actual Web archive or computing a ranking function on internal characteristics. Neither is it required to have the entire archive on-site. Only if the user clicks on a result it opens a version from the queried time period using the Internet Archive's Wayback Machine or any other Memento compliant Web archive\footnote{\url{http://mementoweb.org}}.

By that, \textsf{Tempas} serves as an effective entry point to Web archives and naturally provides high accurate results with respect to the underlying external resource. Researchers using these results in their studies should be aware of the bias introduced by the dataset, however, it allows them to build a corpus for their research which is well-defined and easily comprehensible. While more advanced search and ranking methods are often complex and their performance is questionable, especially on temporal datasets such as Web archives, the results on \textsf{Tempas} solely correspond to their temporal popularity on the external data source. Besides ranking up the most temporally as well as topically related websites, like \textit{change.gov} in our example, it also filters the vast amount of noise and low quality websites on the Web, which are not included or less frequent in the external data. Our demo can be accessed at:\\ \texttt{\url{http://tempas.l3s.uni-hannover.de:8887}}

\section{Tempas Architecture}
\label{sec:architecture}

With \textsf{Tempas} we have built an indexing framework over websites or their URLs respectively, based on tags and timestamps from the \emph{Delicious} dataset (cf. Section~\ref{sec:dataset}) to answer temporal queries (cf. Section~\ref{sec:models}). In addition, to encourage exploration of the collection we rank related documents and suggest similar tags for reformulation and expansion of the query intent (cf. Section~\ref{sec:retrieval}).

\subsection{Dataset}
\label{sec:dataset}

Our work on \textsf{Tempas} is based on the data of \textit{Delicious} from 2003 to 2011, collected by \cite{socialbm0311}. The dataset, called \textit{SocialBM0311}, has been published online and is freely available\footnote{\url{http://www.zubiaga.org/datasets/socialbm0311}}. It contains the complete bookmarking activity for almost 2 million users from the launch of the social bookmarking website in 2003 to the end of March 2011 with 339,897,227 bookmarks, 118,520,382 unique URLs, 14,723,731 unique tags and 1,951,207 users. Its size is 11GB of compressed, tab-separated text data with each line in the following form:

\texttt{<url\_md5   user\_id   url   unix\_timestamp   tags>}

In the following we will refer to a URL as website or use the terms interchangeably. Every record in the dataset with its specific time is referred to as a version of the website. In the final system this is linked to a capture in the archive, which is a snapshot of when the website was crawled.

\subsection{Data and Query Model}
\label{sec:models}

We operate on the tag dataset described above where websites, considered as our documents $d \in \mathcal{D}$, are tagged with labels $l \subseteq \mathcal{L}$ at a given time $t \in \mathcal{T}$. We allow for a discrete representation of time and assume a granularity of days. Each tuple in the dataset can be represented as a triple $(d,l,t) \in \mathcal{D} \times 2^\mathcal{L} \times T$. Note that such a tuple represents the version of the document $d$ at the time $t$. A temporal query $q~=~(q_l\,,\,q_t)$ has a text component $q_l \in \mathcal{L}$ and a time period of focus $q_t \in \mathcal{T} \times \mathcal{T}$. We require that the results for the temporal selection induced by the query return versions of the documents which are valid in $q_t$. In what follows, we use the terms websites and documents interchangeably.

\subsection{Ranking Documents and Tags}
\label{sec:retrieval}

For designing the retrieval model we take the following desiderata into consideration:
\begin{enumerate}
\item Most relevant websites in a given time interval with respect to certain query tags are also most frequently tagged with the query terms during this time frame.
\item More relevant versions of a website in a given time interval with respect to a set of query tags are tagged with more of these tags and less other tags.
\item Most frequently co-occurring tags of given query tags in a certain time interval represent their most related tags/topics during this time frame. 
\end{enumerate}

First, we retrieve a set of relevant documents $R(q)$ which are valid in $q_t$. A document is considered relevant if its versions in $q_t$ cover the query terms $q_l$. In other words, the union of tags of all versions of $d \in R(q)$ in $q_t$ must cover $q_l$. 

For ranking we follow a nested ranking approach in which we first rank documents or websites and then rank its corresponding versions. Based on our desiderata, we compute the score of each document as the product of the \emph{mutual information} of the document $d \in R(q)$ and the query terms in $q_t$ along with the popularity of the document. The popularity of the document $d$ is measured by the frequency of versions of $d$ tagged in $q_t$. Note that there could be multiple tuples for the same document with or without the same tag sets. Next, following the second desiderata, we rank the versions for a given document based on vanilla counts of query tags associated with each version.

Finally, since we are also interested in retrieving related tags, we also retrieve a set of relevant co-occurring tags given $q_l$. A tag is deemed relevant if it co-occurs with the query terms in $q_l$. Similar to the document relevance, a tag might be relevant even if it does not co-occur with all tags in $q_l$ for a version of $d \in R(q)$ if it co-occurs with the remainder of the tags is some other version of $d$. The tags are scored and aggregated across all documents based on weighted counts of their co-occurrences to give a final ranked list of most relevant co-occurring tags. 

\subsection{Index Structures}
\label{sec:indexing}

The core of \textsf{Tempas} is a collection of indexes and mappings, which are tailored for retrieving the above described result sets (s. Sec.~\ref{sec:retrieval}). All of them are built to provide retrieval with a monthly granularity. We created indexes to retrieve tags as well as websites based on a query consisting of tags $q_l$ for a time period $q_t$ (i.e., \textit{TagTagMapping}, \textit{TagUrlMapping}). Furthermore, we created a year and monthly based index to retrieve tags without providing tags as input (i.e., \textit{YearTagMapping}, \textit{MonthTagMapping}) for exploratory search for a particular time interval without providing tags as input. Another index allows retrieving all versions of a website that have been tagged during a given time period together with the tags (i.e., \textit{UrlTagMapping}). For a compact index structure, two mappings assign ids to tags and websites, which are used exclusively in all other indexes and mappings (i.e., \textit{IdTagMapping}, \textit{IdUrlMapping}). Even though there is much room for improvement and optimization of temporal indexes \citep{anand_index_2012, anand_efficient_2010}, the rather simple mappings, which can easily be constructed using a distributed data processing platform like \emph{Hadoop}, already go a long way. 








All indexes are ordered by their keys (i.e., ID, year, month or pair of tag and year or website and year, respectively). The fetching of the indexes is realized by web services, which are invoked separately for tags and websites as well as a single fetch for the versions with tags of each website. After fetching the data, all items (i.e., tags, websites or versions) are ordered according the ranking functions in Sec.~\ref{sec:retrieval}.

The query with tags as well as the time period, consisting of start year and month as well as end year and month, can be entered and selected at the top of the page, using the search input and the three sliders as shown in Fig.~\ref{fig:screenshot}.

\section{Conclusions and Outlook}
\label{sec:conclusions}

Tags have been proven to be a reasonable surrogate for the actual content of a website. As indexing complete Web archives in a full text index as well as providing temporal IR capabilities to this appears to be a big challenge, meta information, such as tags from social bookmarking services, are a good way to mimic the real content of different versions of a website. We have not yet evaluated the effectiveness of our method in terms of meeting a users information need, however, it is definitely a great improvement over accessing a Web archive by providing the exact URL and time of a website's version, like most Web archives only allow today.

The different levels of co-occurring tags to be considered as sub-topics of the original query in \textsf{Tempas} provide an easy way to explore the available dataset. The ability to specify a time period to search in enables temporal IR capabilities for Web archives. However, in order to present the retrieved websites to the user, in its current state, \textsf{Tempas} only passively requests the intended version of a website from the Internet Archive's Wayback Machine (cf. Sec.~\ref{sec:architecture}). Therefore, we do not know whether an archived version is actually available and we do not have access to data like the title of the website, which therefore cannot be included in the search results. In the context of the ALEXANDRIA project\footnote{\url{http://alexandria-project.eu}} we own a subset of this archive, comprising the complete archived part of the German Web, which we have full access to. In coming versions of \textsf{Tempas} we will integrate this to provide richer information to the user. It will also allow us to compute statistics of the retrieved result set, such as the fraction of the actually archived Web that is covered by \textsf{Tempas} as well as the time gaps between tagged and archived versions of the websites.

In future research we are going to look into more sophisticated retrieval as well as ranking models. Also different visualizations and ways to explore archives are subject for further investigation. While we have shown how tags can serve as surrogates for the content of a website in temporal IR, there are other IR related challenges that need to be tackled in the future in order to enable the same convenience for temporal Web archive search that we are used to in common search engines on the current Web. For instance, query suggestions and reformulations, which support the users in formulating their information needs, are not available for archives yet. This is mainly due to the temporal characteristic, which raises new challenges for theses tasks, as well as the lack of query logs, which are commonly used for this purpose. We will investigate how tags from social bookmarking services can serve as surrogates for query logs as well as other required data in temporal IR on Web archives as well. Furthermore, we are planning on incorporating different meta data than tags that is available in a temporal fashion or can be extracted from an archive as derivative dataset. This might include postings on social websites such as Twitter as well as host graphs of websites, which can be derived from the crawls of archived websites.

\bibliographystyle{unsrtnat}
\bibliography{references}

\end{document}